\documentclass[twocolumn]{aastex631}
\hypersetup{linkcolor=red,citecolor=blue,filecolor=cyan,urlcolor=blue}
\usepackage{enumitem}

\begin{document}

\title{The Binary Fraction of B-type Runaway Stars from LAMOST DR8}

\author[0009-0002-2432-5603]{Kun Chen}
\affiliation{International Centre of Supernovae (ICESUN), Yunnan Key Laboratory of Supernova Research, Yunnan Observatories, Chinese Academy of Sciences (CAS), Kunming 650216, People's Republic of China}
\affiliation{School of Astronomy and Space Science, University of Chinese Academy of Sciences (UCAS), Beijing 100049, People's Republic of China}

\author[0000-0001-9989-9834]{Yanjun Guo}
\affiliation{International Centre of Supernovae (ICESUN), Yunnan Key Laboratory of Supernova Research, Yunnan Observatories, Chinese Academy of Sciences (CAS), Kunming 650216, People's Republic of China}

\author[0000-0003-4265-7783]{Dengkai Jiang}
\affiliation{International Centre of Supernovae (ICESUN), Yunnan Key Laboratory of Supernova Research, Yunnan Observatories, Chinese Academy of Sciences (CAS), Kunming 650216, People's Republic of China}

\author[0000-0001-9204-7778]{Zhanwen Han}
\affiliation{International Centre of Supernovae (ICESUN), Yunnan Key Laboratory of Supernova Research, Yunnan Observatories, Chinese Academy of Sciences (CAS), Kunming 650216, People's Republic of China}

\author[0000-0001-5284-8001]{Xuefei Chen}
\affiliation{International Centre of Supernovae (ICESUN), Yunnan Key Laboratory of Supernova Research, Yunnan Observatories, Chinese Academy of Sciences (CAS), Kunming 650216, People's Republic of China}

\correspondingauthor{Kun Chen, Yanjun Guo, Dengkai Jiang, Zhanwen Han, Xuefei Chen}
\email{chenkun@ynao.ac.cn, guoyanjun@ynao.ac.cn, \\
dengkai@ynao.ac.cn, zhanwenhan@ynao.ac.cn, cxf@ynao.ac.cn}

%% Note that the \and command from previous versions of AASTeX is now
%% depreciated in this version as it is no longer necessary. AASTeX 
%% automatically takes care of all commas and "and"s between authors names.

%% AASTeX 6.31 has the new \collaboration and \nocollaboration commands to
%% provide the collaboration status of a group of authors. These commands 
%% can be used either before or after the list of corresponding authors. The
%% argument for \collaboration is the collaboration identifier. Authors are
%% encouraged to surround collaboration identifiers with ()s. The 
%% \nocollaboration command takes no argument and exists to indicate that
%% the nearby authors are not part of surrounding collaborations.

%% Mark off the abstract in the ``abstract'' environment. 

\begin{abstract}

Runaway stars are defined as stars that depart from their birth clusters at high peculiar velocities. There are two main mechanisms for the formation of runaway stars, i.e., the binary-supernova scenario (BSS) and the dynamical ejection scenario (DES). Investigating the binary fraction of runaway stars is an important step in further exploring the relative significance of the two mechanisms. We analyzed the binary fraction of 203 Galactic B-type runaway stars identified in the Large Sky Area Multi-Object Fiber Spectroscopic Telescope Data Release 8 database. Our analysis of radial velocity variations in the runaway star sample reveals an observed spectroscopic binary fraction of $5.4\%\pm 1.6\%$, representing the proportion of objects that exhibit statistically significant variations in radial velocity with amplitudes larger than $\rm 16~km~s^{-1}$. We employed a Monte Carlo method to correct for observational biases and determined an intrinsic binary fraction of $27\%\pm 8\%$. The period and mass ratio distributions that best reproduce the observation are $f(P)\propto P^{-5.7}$ for $1\leq P\leq 1000$ days, and $f(q)\propto q^{-3.6}$ for $0.1\leq q\leq 1.0$, indicating a preference for binaries with shorter periods and less massive companions compared to a uniform distribution. The intrinsic binary fraction, in combination with previous studies on the binary fractions of runaway stars formed by the BSS and the DES, implies that both scenarios contribute comparably to the formation of Galactic B-type runaway stars, where the ratio of the BSS to the DES is 0.86.

\end{abstract}

%% Keywords should appear after the \end{abstract} command. 
%% The AAS Journals now uses Unified Astronomy Thesaurus concepts:
%% https://astrothesaurus.org
%% You will be asked to selected these concepts during the submission process
%% but this old "keyword" functionality is maintained in case authors want
%% to include these concepts in their preprints.


%% From the front matter, we move on to the body of the paper.
%% Sections are demarcated by \section and \subsection, respectively.
%% Observe the use of the LaTeX \label
%% command after the \subsection to give a symbolic KEY to the
%% subsection for cross-referencing in a \ref command.
%% You can use LaTeX's \ref and \label commands to keep track of
%% cross-references to sections, equations, tables, and figures.
%% That way, if you change the order of any elements, LaTeX will
%% automatically renumber them.
%%
%% We recommend that authors also use the natbib \citep
%% and \citet commands to identify citations.  The citations are
%% tied to the reference list via symbolic KEYs. The KEY corresponds
%% to the KEY in the \bibitem in the reference list below. 

\section{Introduction} \label{sec:intro}

Runaway stars are those that rapidly escape from their parent clusters or associations, with peculiar velocities exceeding $\rm \sim 30-40~km~s^{-1}$ \citep[e.g.,][]{Blaauw+Morgan.1954.ApJ,Blaauw.1956.PASP,Gies+Bolton.1986.ApJS,De_Donder+.1997.A&A,Hoogerwerf+.2000.ApJ,de_Wit+.2005.A&A,Dray+.2005.MNRAS,Eldridge+.2011.MNRAS,Boubert+Evans.2018.MNRAS}. The majority of the observed runaway stars are massive early-type stars \citep[e.g.,][]{Blaauw.1961.BAN,Gies.1987.ApJS,Hoogerwerf+.2001.A&A,Mdzinarishvili+Chargeishvili.2005.A&A}. The population of early-type runaway stars exhibits a velocity dispersion of $\rm \sim 30~km~s^{-1}$ \citep[e.g.,][]{Stone.1991.AJ,Tetzlaff+.2011.MNRAS,Carretero-Castrillo+.2024.BSRSL}, greater than the $\rm \sim 10~km~s^{-1}$ velocity dispersion observed in early-type stars within young star clusters. Additionally, the binary fraction of early-type runaway stars is relatively low \citep[e.g.,][]{Gies.1987.ApJS,Mason+.1998.AJ,Mason+.2009.AJ,Aldoretta+.2015.AJ} compared to that of early-type stars in young star clusters, which have a binary fraction $>50\%$ \citep[e.g.,][]{Sana+.2012.Sci,Sana+.2013.A&A,Dunstall+.2015.A&A,Guo+.2022.A&A,Guo+.2022.RAA,Chen+.2024.PrPNP}.

Currently, there are two primary formation scenarios for early-type runaway stars. One is the binary-supernova scenario (BSS), stemming from a suggestion by \cite{Zwicky.1957.moas.book}; the other is the dynamical ejection scenario (DES), first proposed by \cite{Poveda+.1967.BOTT}. In the BSS, the primary star of massive binary system will undergo a core-collapse supernova (SN) first and remain a compact object, either a neutron star or a black hole. If the SN explosion ejects more than half of the total system mass or imparts a significant natal kick to the compact object, the less massive companion star will be ejected from the compact object and is likely to become a single runaway, with a space velocity on the order of its final orbital velocity before the SN \citep{Blaauw.1961.BAN,Leonard+Duncan.1988.AJ,Renzo+.2019.A&A}. Otherwise, the companion star and the compact object will remain bound and may possibly form a binary runaway, which could be detected as a high-mass X-ray binary (HMXB) \citep[e.g.,][]{Gott.1971.Natur,van_den_Heuvel+.2000.A&A}. In the DES, early-type runaway stars are ejected from dense, compact clusters via gravitational interactions among massive stars \citep[e.g.,][]{Poveda+.1967.BOTT,Leonard.1991.AJ}. Two possible ways to form early-type runaway stars are binary--single star encounters \citep[e.g.,][]{Hut.1983.ApJ,Hut.1993.ApJ,Hut+Bahcall.1983.ApJ} and binary--binary interactions \citep[e.g.,][]{Mikkola.1983.MNRAS,Mikkola.1984.MNRAS.a,Mikkola.1984.MNRAS.b}. \cite{Hoffer.1983.AJ} suggested that the latter is the most efficient interaction to produce early-type runaway stars. In most cases, the outcome of a binary--binary interaction will be two single stars and one binary system \citep{Hoffer.1983.AJ,Mikkola.1983.MNRAS}. Since the binary system is the most massive of the three interaction products, it is unlikely to acquire a large system velocity, i.e., only a few of these binary systems will gain sufficient velocity to become binary runaways \citep{Hoogerwerf+.2001.A&A}. Therefore, the binary fraction of early-type runaway stars which are formed via the DES is expected to be low \citep[e.g.,][]{Dorigo_Jones+.2020.ApJ}.

For decades, many studies have been conducted on the binary fraction of early-type runaway stars. The most investigated binary property is the observed fraction of spectroscopic binaries. \cite{Mason+.2009.AJ} showed that 12 out of 42 Galactic O-type runaway stars are spectroscopic binaries, corresponding to an observed fraction of $29\%$. \cite{Chini+.2012.MNRAS} found a spectroscopic binary fraction of $69\%$ for 54 Galactic O-type runaway stars. Another study claimed an observed binary fraction of $28\%$ among the 29 Galactic O-type runaway stars investigated through multiepoch spectroscopy \citep{Aldoretta+.2015.AJ}. 

In addition to the observed fraction of spectroscopic binaries, the binary fractions of runaway stars formed through the BSS and the DES can be studied using binary population synthesis and $N$-body simulations, respectively. For the BSS, \cite{Portegies_Zwart.2000.ApJ} carried out binary population synthesis calculations to explore the origin of the BSS O-type and B-type runaway stars with velocities larger than $\rm 25~km~s^{-1}$. They obtained a binary fraction of $20\%$ for the BSS B-type runaway stars and $40\%$ for the BSS O-type runaway stars. For the DES, \cite{Perets+Subr.2012.ApJ} made use of $N$-body simulations of massive clusters to explore the theoretical distribution of ejection velocities for runaway stars. Their studies predicted that about $33\%$ of the DES O-type runaway stars, with ejection velocities larger than $\rm 30~km~s^{-1}$, were binaries. \cite{Oh+.2015.ApJ} also estimated the binary fraction by conducting direct $N$-body simulations. Their most realistic models yielded a binarity of $\sim 19\%-29\%$ for the DES O-type runaway stars with velocities greater than $\rm 30~km~s^{-1}$. 

Until now, the relative importance of two formation scenarios has not been well established. \cite{Dorigo_Jones+.2020.ApJ} studied the kinematics of 304 field OB stars within the Small Magellanic Cloud (SMC) to constrain the ratio of the DES to the BSS runaway stars with local residual transverse velocities larger than $\rm 30~km~s^{-1}$, which is approximately $2-3$. \cite{Phillips+.2024.ApJ} confirmed that the DES dominates over the BSS by a factor of $\sim 1.7$ for runaway stars with space velocities greater than $\rm 30~km~s^{-1}$, based on the kinematic analysis of 336 field OB and OBe stars in the SMC. All of the above studies favor the dominance of the DES in runaway star formation, while others suggest that the BSS may play a more important role than the DES. \cite{Hoogerwerf+.2001.A&A} found that roughly $2/3$ and $1/3$ of the runaway stars, for which the origin was confirmed, are produced by the BSS and the DES, respectively. \cite{Sana+.2022.A&A} obtained a detection ratio close to $2:1$ between a slowly moving but rapidly rotating runaway population and a fast-moving but slowly spinning one, suggesting that the BSS dominates the massive runaway population within 30 Doradus.

One way to determine the relative importance is to study the statistical properties of a complete runaway star sample \citep{Hoogerwerf+.2001.A&A}. The binarity of runaway stars is one of the most important statistical properties. Due to the strategy of observations, the signal-to-noise ratio (S/N) of the data, and the inclination of binary systems, spectroscopy may not be able to detect all the binaries. Thus, the observed fraction of spectroscopic binaries is actually a lower limit on the true binary fraction. If we can obtain an accurate true binary fraction for runaway stars, it would greatly support further research on the relative importance of two scenarios. 

Previous studies have proposed a Monte Carlo method for investigating the true binary fractions of various early-type star samples based on multiepoch spectroscopy \citep{Sana+.2013.A&A,Dunstall+.2015.A&A,Guo+.2022.A&A,Guo+.2022.RAA}. According to their methodology, here we try to constrain the true binary fraction of a Galactic B-type runaway star sample, as published by \cite{Guo+.2024.ApJS}. 

In Section \ref{sec:data}, we provide a brief introduction to the Galactic B-type runaway star sample and the radial velocity (RV) data. Section \ref{sec:analysis} describes the method for identifying spectroscopic binaries and reports the observed spectroscopic binary fraction of the runaway star sample. Our work on constraining the intrinsic multiplicity properties of the runaway star sample is presented in Section \ref{sec:properties}, and we discuss the results in Section \ref{sec:discussion}. The conclusions of our work are summarized in Section \ref{sec:conclusion}.

\begin{figure}
\plotone{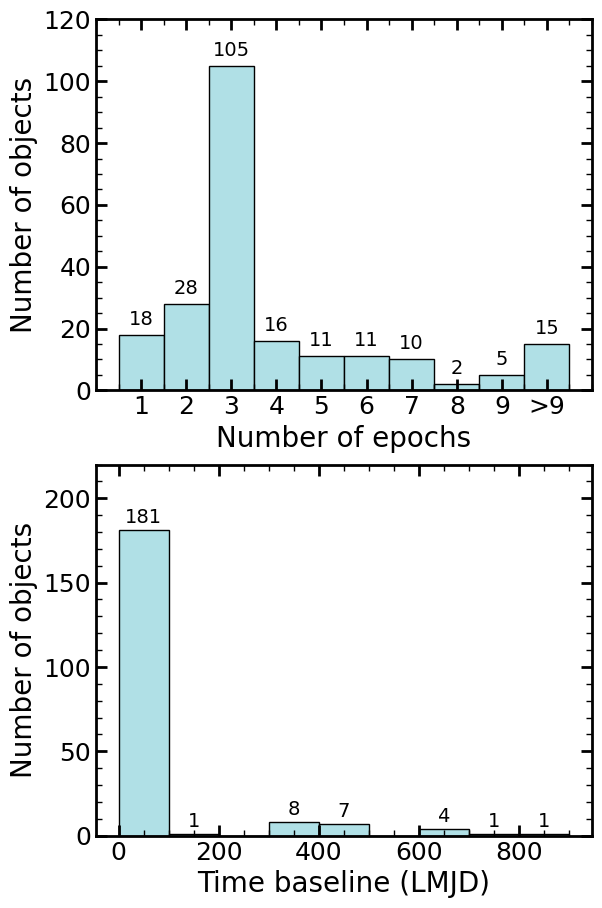}
\caption{\textit{Top panel}: The number distribution of observations for stars in the B-type runaway star sample. This indicates that 18 stars have been observed only once, $48\%$ of the stars have been observed three times, and $32\%$ have been observed at least four times. \textit{Bottom panel}: The number distribution of time intervals between the first and last observation for each star. The 18 stars with a single observation are excluded. This shows that $89\%$ of the 203 stars with at least two observations have a time baseline of less than 100 days.
\label{fig:observation}}
\end{figure}

\section{Runaway star sample and RV data} \label{sec:data}

The Large Sky Area Multi-Object Fiber Spectroscopic Telescope (LAMOST) is a 4 m quasi-meridian reflecting Schmidt spectroscopic survey telescope located at the Xinglong Station of the National Astronomical Observatories, Chinese Academy of Sciences. The telescope is equipped with 4000 optical fibers, enabling simultaneous spectroscopic observations of thousands of celestial objects in a single exposure. In addition, it features both low-resolution spectrographs (with a resolving power of $R\sim 1800$) and medium-resolution spectrographs ($R\sim 7500$), making it well suited for a broad range of astrophysical studies \citep{Cui+.2012.RAA,Deng+.2012.RAA,Zhao+.2012.RAA}.

In this paper, we analyze the sample of Galactic early-type runaway stars identified from the LAMOST Data Release 8 (DR8) database by \cite{Guo+.2024.ApJS}. In their work, the authors adopted a runaway criterion of peculiar space velocity $\rm >43~km~s^{-1}$ to identify runaway stars. Based on RV data from LAMOST DR8 and astrometric solutions from Gaia Data Release 3, they identified 229 runaway candidates among a sample of 4432 Galactic early-type stars. According to the effective temperatures reported in their study, the early-type runaway star sample is dominated by B-type stars, with a few O-type stars. To construct a pure B-type star sample for homogeneous analysis, we excluded eight O-type stars with effective temperatures above $\rm 30,000~K$ from the runaway star sample. To obtain the RVs of the 221 B-type runaway stars, we cross matched the sample with the RV database updated using the self-consistent method provided by \cite{Zhang+.2021.ApJS}.

In Table \ref{tab:catalogs}, we list the LAMOST observation ID (OBSID) for each star, along with its equatorial coordinates (R.A. and Decl.), Local Modified Julian Date (LMJD), S/N, measured RV, associated observational uncertainty ($\rm \sigma_{RV}$), and group index (GroupID). Data sets sharing the same GroupID correspond to multiple observations of the same star. Figure \ref{fig:observation} displays the statistics of the number of observations and the observational baselines for stars in the B-type runaway star sample. There are 18 stars with only a single observation. To perform the binarity analysis in the following section, at least one pair of RV measurements is required for each star. So, we adopted 203 B-type runaway stars with at least two observations as the final sample for the further analysis.

\begin{deluxetable*}{lccccccc}
% \tabletypesize{\scriptsize}
\tablewidth{0pt} 
\tablecaption{The RVs of the Runaway Star Sample \label{tab:catalogs}}
\tablehead{
\colhead{OBSID} & \colhead{R.A.}    & \colhead{Decl.}   & \colhead{LMJD}  & \colhead{S/N}  & 
\colhead{RV}                      & \colhead{$\sigma_{\rm RV}$}       & \colhead{GroupID} 
\\
\colhead{}      & \colhead{(deg)} & \colhead{(deg)} & \colhead{(day)} & \colhead{}     & 
\colhead{($\rm km~s^{-1}$)}       & \colhead{($\rm km~s^{-1}$)}       & \colhead{}
} 
\startdata 
609207171 & 0.13011 & 54.41446 & 58088.80972 & 52.44  & 27.75004   & 1.93704 & 1 \\
609207171 & 0.13011 & 54.41446 & 58088.79097 & 55.36  & 28.40188   & 1.89850 & 1 \\ 
609207171 & 0.13011 & 54.41446 & 58088.80069 & 56.26  & 32.23582   & 1.71857 & 1 \\
609213158 & 1.18358 & 57.46465 & 58088.80069 & 103.91 & 52.21255   & 0.88252 & 2 \\ 
609213158 & 1.18358 & 57.46465 & 58088.80972 & 104.57 & 52.21972   & 0.74779 & 2 \\ 
609213158 & 1.18358 & 57.46465 & 58088.79097 & 101.95 & 53.99710   & 0.48414 & 2 \\ 
694509085 & 1.94221 & 11.15507 & 58449.79653 & 138.46 & -269.29089 & 0.31843 & 3 \\ 
694509085 & 1.94221 & 11.15507 & 58449.78056 & 149.89 & -269.38200 & 0.17410 & 3 \\ 
\enddata
\tablecomments{Only a portion of this table is shown here to demonstrate its form and content. A machine-readable version of the full table is available.}
\end{deluxetable*}

\section{Multiplicity analysis} \label{sec:analysis}

\subsection{Criteria for binarity} \label{subsec:criteria}

To identify the spectroscopic binaries in our sample, we adopted binarity criteria similar to those used by \cite{Sana+.2013.A&A}, where an object is classified as a spectroscopic binary if at least one RV pair simultaneously meets
\begin{equation}
\frac{|v_i-v_j|}{\sqrt{\sigma_i^2+\sigma_j^2}}>4.0 \quad \text{and} \quad |v_i-v_j|>C,
\label{equ:criteria}
\end{equation}
where $v_{i}$ is the RV estimated from the spectrum at epoch $i$, and $\sigma_{i}$ is its associated measurement error. To verify whether the adopted significance threshold effectively limits the number of false positives, we conducted Monte Carlo simulations. The simulations suggested that the confidence threshold of 4.0 indicated a probability of finding about one false detection in every 1090 objects, given the sample size, the number of RV pairs, and the uncertainties of the RV measurements.

\begin{figure}
\plotone{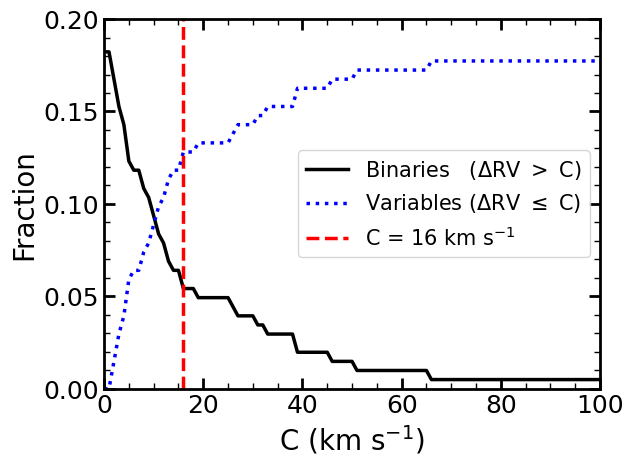}
\caption{The variation in the fraction of spectroscopic binaries and RV variables as a function of the RV variation threshold $C$. The red vertical dashed line represents the adopted threshold $C=\rm 16~km~s^{-1}$.
\label{fig:threshold}}
\end{figure}

The minimum amplitude threshold $C$ of RV variation is set to reject false detections of single stars with atmospheric activity or pulsations. We consider an object as a spectroscopic binary only if it meets both criteria of Equation (\ref{equ:criteria}) simultaneously. Those satisfying the first criterion of Equation (\ref{equ:criteria}) but failing the second are regarded as RV variables. The fraction of spectroscopic binaries and RV variables in our sample for different threshold $C$ is shown in Figure \ref{fig:threshold}. This presents a clear kink at about $\rm 16~km~s^{-1}$. We suggest that below the kink, the rapid decrease in the fraction of binaries as the threshold $C$ increases indicates that the criteria filter out both binaries and RV variables. And above the kink, the sample is probably dominated by binaries, leading to a more gradual change in the fraction. So, the kink can be used to determine the threshold $C$ value. \cite{Sana+.2013.A&A} and \cite{Dunstall+.2015.A&A} applied the same approach to set the threshold $C$ to $\rm 20~km~s^{-1}$ for O-type stars and $\rm 16~km~s^{-1}$ for B-type stars, respectively. Since our sample consists exclusively of B-type stars, we adopted $C=\rm 16~km~s^{-1}$ as the final threshold.

\subsection{Observed binary fraction} \label{subsec:OBF}

According to Equation (\ref{equ:criteria}) with $C=\rm 16~km~s^{-1}$, we found 11 spectroscopic binaries from 203 B-type runaway stars that have at least one pair of RV measurements. The observed spectroscopic binary fraction is $f_{\rm bin}^{\rm obs}=5.4\%\pm 1.6\%$, where the $1\sigma$ uncertainty was estimated using a bootstrap analysis following the method of \cite{Raghavan+.2010.ApJS}. Furthermore, a total of 26 stars meet the first criterion but fell below the $\Delta \rm RV$ cutoff, and are therefore classified as RV variables. They comprise $12.8\%\pm 2.3\%$ of the total sample. These objects might be long-period binaries with small RV variations, short-period binaries with small phase differences of spectroscopic observations, or single stars with atmospheric variability. Hence, the observed fraction of spectroscopic binaries here is merely a lower limit on the true binary fraction of the sample. In the next section, we will employ a Monte Carlo approach to correct for observational biases and investigate the true binary fraction.

\begin{figure*}
\plotone{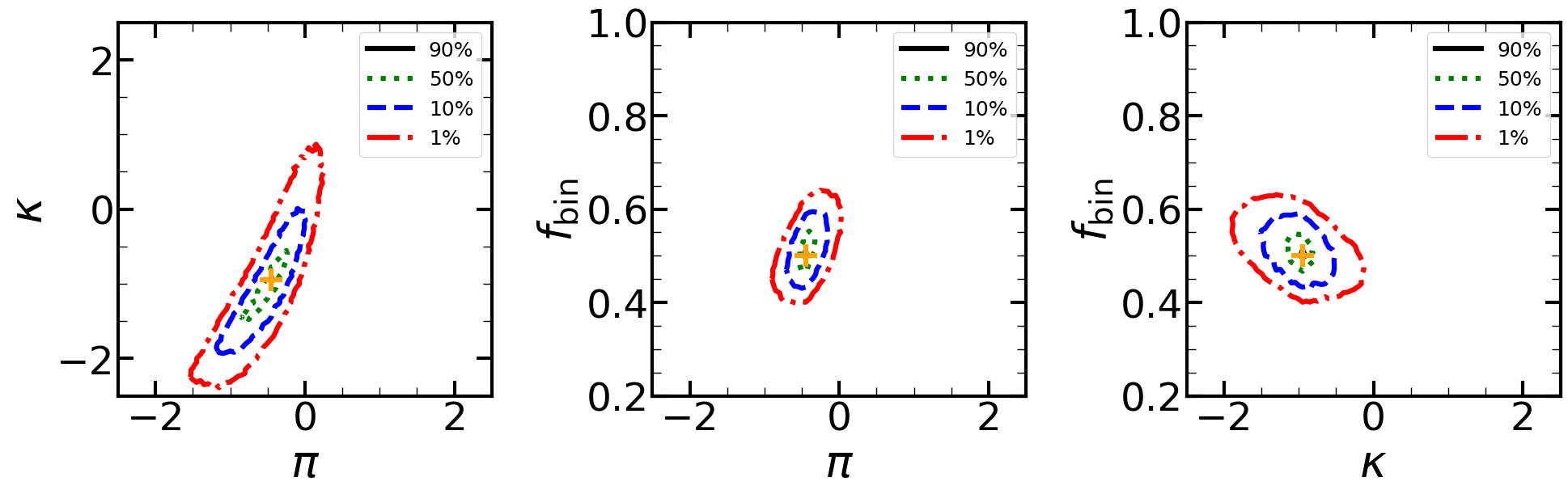}
\caption{Projections of the GMF onto planes defined by the various pairs of $\pi$, $\kappa$, and $f_{\rm bin}$, as obtained from analyzing the data of \cite{Sana+.2013.A&A}. The orange cross (+) indicates the location of the absolute maximum. The red dashed--dotted, blue dashed, green dotted, and black solid contours represent the loci of equal-values corresponding to $1\%$, $10\%$, $50\%$, and $90\%$ of the absolute maximum, respectively.
\label{fig:validation}}
\end{figure*}

\section{Intrinsic multiplicity properties} \label{sec:properties}

As mentioned in the previous section, the number of detected binaries actually represents a lower-limit estimate of the true binary count in the runaway star sample. In addition, the binary detection probability is influenced by the cadence of LAMOST observations, the accuracy of RV measurements, and the intrinsic distribution of orbital parameters. In order to correct for these biases and to constrain the true binary fraction and the distributions of orbital parameters, we constructed a Monte Carlo simulation based on an approach similar to that of \cite{Sana+.2013.A&A}.

\subsection{Monte Carlo method} \label{subsec:method}

In the Monte Carlo approach, we modeled numerous stellar populations, where the orbital parameters of the binaries were randomly drawn from specified parent distributions. In the stellar populations, a Bernoulli trial with a success probability of $f_{\rm bin}$ (the intrinsic binary fraction) was performed for each star. If the result of the trial is 1, the star will be flagged as a binary; otherwise, it will be flagged as a single star. For each binary in the simulated stellar populations, we calculated the RVs of its primary star at the times of our observations, and randomly introduced measurement uncertainties using a Gaussian distribution with a standard deviation equal to the associated observational errors. For each simulated single star, its RVs at the observational epochs were randomly sampled from a Gaussian distribution with a fixed mean (we adopted 0 as the mean value here) and a standard deviation determined by the corresponding observational errors.
 
Next, we applied the binary detection criteria of Equation (\ref{equ:criteria}) to the synthetic stellar populations to obtain observational quantities. For each combination of the specified parent distributions, we repeated the above process 100 times and used the following quantities, acquired from all the 100 simulations, as the final simulated observational features: (i) the simulated observed fraction, $f_{\rm bin}^{\rm sim}$, of spectroscopic binaries; (ii) the distribution of the maximum amplitude, $\Delta \rm RV$, of significant RV variations for each binary identified via Equation (\ref{equ:criteria}); and (iii) the distribution of the minimum observational time interval, $\Delta \rm LMJD$, for the RV pairs with significant variations of each detected binary. 

Finally, we computed the Kolmogorov--Smirnov probabilities between the observed empirical cumulative distribution functions (ECDFs) for $\Delta \rm RV$ and $\Delta \rm LMJD$ and their simulated ones from the 100 simulations. Moreover, the binomial probability $B(N_{\rm bin},N,f_{\rm bin}^{\rm sim})$ was also calculated to evaluate the likelihood of detecting the same number of binaries ($N_{\rm bin}$) as identified in the runaway star sample, considering the sample size $N$ and the simulated observed binary fraction $f_{\rm bin}^{\rm sim}$. To evaluate how well the simulated sample matches the observed one, we follow the approach of \cite{Sana+.2013.A&A} by adopting the product of the three quantities above as the global merit function (GMF):
\begin{equation}
\resizebox{0.9\columnwidth}{!}{$\text{GMF}=P_{\rm KS}(\Delta \text{RV})\times P_{\rm KS}(\Delta \text{LMJD})\times B(N_{\rm {bin}},N,f_{\rm {bin}}^{\rm {sim}}).$}
\label{equ:GMF}
\end{equation}

Since the GMF is the product of three probabilities, it indicates that the simulated sample best matches the observed one when the GMF reaches its maximum value. In addition, as we set the intrinsic binary fraction and parameter distributions of the simulated sample, the simulated sample corresponding to the maximum value of the GMF can be used to constrain the true binary fraction and the orbital parameter distributions of the runaway star sample. 

\begin{deluxetable*}{lcccccc}
% \tabletypesize{\scriptsize}
\tablewidth{0pt} 
\tablecaption{Properties of the Parameters and Variables Used in our Monte Carlo Simulation \label{tab:parameters}}
\tablehead{
\colhead{Parameter} & \colhead{PDF} & \colhead{Domain} & \colhead{Unit} & \colhead{Variable} & 
\colhead{Range}     & \colhead{Step}
} 
\startdata 
$M_1$         & $M_1^{-2.3}$    & 2 - 16          & $M_{\odot}$ & ···           & ···           & ···  \\ 
$P$           & $P^{\pi}$       & 1 - 1000        & day         & $\pi$         & -11.0 to -1.0 & 0.1  \\
$q$           & $q^{\kappa}$    & 0.1 - 1.0       & ···         & $\kappa$      & -6.0 to -1.0  & 0.1  \\
$e$           & $e^{-0.5}$      & $10^{-5}$ - 0.9 & ···         & ···           & ···           & ···  \\
$T_0$         & Uniform         & 0 - $P$         & day         & ···           & ···           & ···  \\
$\omega$      & Uniform         & 0 - 2$\pi$      & rad         & ···           & ···           & ···  \\
$i$           & $\text{sin}(i)$ & 0 - $\pi$/2     & rad         & ···           & ···           & ···  \\
$f_{\rm bin}$ & ···             & ···             & ···         & $f_{\rm bin}$ & 0.15 - 0.35   & 0.02
\enddata
\tablecomments{The first four columns list the physical parameters, their associated PDFs, their applicable domains, and the units. The last three columns present the free variables, their allowed ranges, and step sizes. The first row provides information on the masses of the primary stars ($M_1$) in the simulated binary systems. The last row displays the investigated range and step size for the intrinsic binary fraction ($f_{\rm bin}$).}
\end{deluxetable*}

\subsection{Validation} \label{subsec:validation}

Based on the procedure described above, we wrote a Python program and applied it to analyze the sample of 360 O-type stars collected from \cite{Sana+.2013.A&A} for code validation. The parent distributions of parameters for the binaries were set as follows: 

\begin{enumerate}[left=0pt]
\renewcommand{\labelenumi}{\roman{enumi}.}

\item The mass of each primary star ($M_1$) is randomly drawn from the initial mass function \citep{Kroupa.2001.MNRAS} over an interval from 15 to 80 $M_{\odot}$, as adopted in \cite{Sana+.2013.A&A}; 

\item In accordance with the prescription of \cite{Sana+.2013.A&A}, orbital periods ($P$), mass ratios ($q=M_2/M_1$), and orbital eccentricities ($e$) follow the power-law distributions: $f(\text{log}_{10}\, P)\propto (\text{log}_{10}\, P)^{\pi}$, $f(q)\propto q^{\kappa}$ and $f(e)\propto e^{\eta}$ over the ranges of $0.15\leq \text{log}_{10}\, P\leq 3.50$, $0.1\leq q\leq 1.0$ and $10^{-5}\leq e\leq 0.9$, respectively. The exponents $\pi$, $\kappa$, and the intrinsic binary fraction $f_{\rm bin}$ of synthetic stellar populations are treated as free variables to construct binary samples with different distributions. Due to the insensitivity of the GMF to eccentricity, we set the exponent $\eta$ to $-0.5$, following \cite{Sana+.2013.A&A};

\item The epoch of periastron ($T_0$) and argument of periapsis ($\omega$) satisfy uniform distributions with the respective ranges of $0\leq T_0\leq P$ and $0\leq \omega\leq 2\pi$. The orbital inclination ($i$) meets a probability density function (PDF) of $\text{sin}(i)$ within the range from 0 to $\pi /2$. The units for $P$ and $T_0$ are set to days.

\end{enumerate}

According to the grid of free variables $[\pi,\kappa,f_{\rm bin}]$ in \cite{Sana+.2013.A&A}, we obtained the maximum value of the GMF at $\pi=-0.45\pm 0.30$, $\kappa=-0.95\pm 0.31$ and $f_{\rm bin}=0.50\pm 0.03$. Compared to the findings of \cite{Sana+.2013.A&A}, where $\pi=-0.45\pm 0.30$, $\kappa=-1.00\pm 0.40$ and $f_{\rm bin}=0.51\pm 0.04$, our results are consistent with theirs given the uncertainties of the optimal variables. The projections of the GMF onto planes defined by the different variable pairs, derived from the analysis of the data from \cite{Sana+.2013.A&A}, are depicted in Figure \ref{fig:validation}.

\subsection{Results} \label{subsec:results}

Before applying our code to analyze the runaway star sample, we made some adjustments to the parent distributions of parameters considering the properties of our sample. Since the runaway star sample is composed entirely of B-type stars, we adjusted the mass distribution range of the primary stars to between 2 and 16 $M_{\odot}$. The distribution of orbital period was modified to a linear form of $f(P)\propto P^{\pi}$ to enhance sensitivity toward short-period binaries \citep{Guo+.2022.RAA}. We also adjusted the orbital period distribution to cover the range from 1 to 1000 days, based on the expected distribution of orbital periods and space velocities of post-SN binaries, as presented in \cite{Portegies_Zwart.2000.ApJ}. The other parameters were set in the same way as stated in Section \ref{subsec:validation}. In Table \ref{tab:parameters}, we list the distributions of parameters with their domains and the grid of free variables. 

\begin{figure*}
\plotone{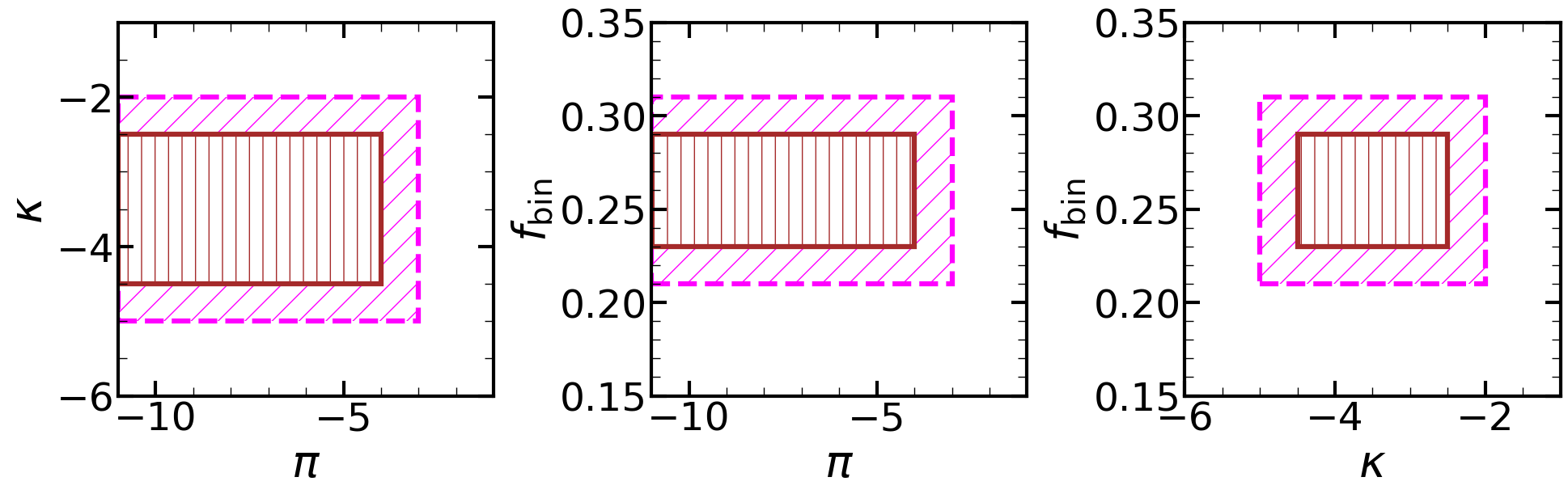}
\caption{Projections of the regions of grid points with different calculation frequencies. The GMF at each gird point within the blank region, magenta--hatched shaded region, and brown vertical--striped shaded region was calculated 100, 500, and 1400 times, in turn, with the final GMF value being taken as the average.
\label{fig:boundary}}
\end{figure*}

\begin{figure*}
\plotone{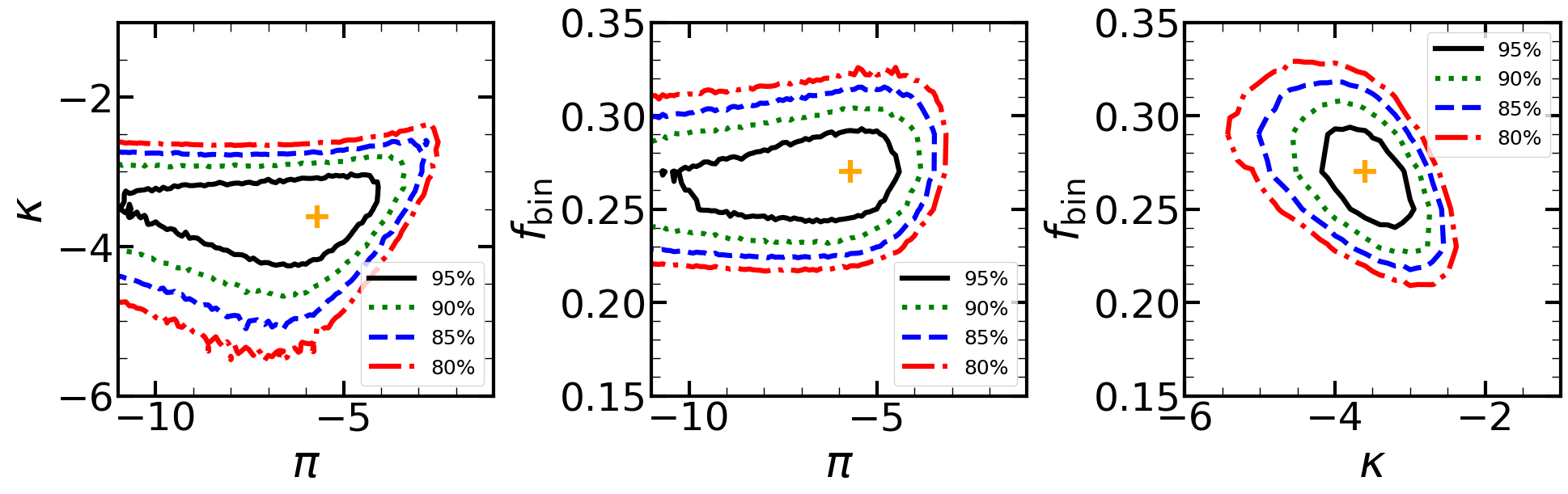}
\caption{Projections of the GMF derived from the analysis of the runaway star sample. The absolute maximum is indicated by the orange cross (+). The loci of equal-values at $80\%$, $85\%$, $90\%$, and $95\%$ of the absolute maximum are displayed by the red dashed--dotted, blue dashed, green dotted, and black solid contours, respectively.
\label{fig:result}}
\end{figure*}

\begin{figure*}
\plotone{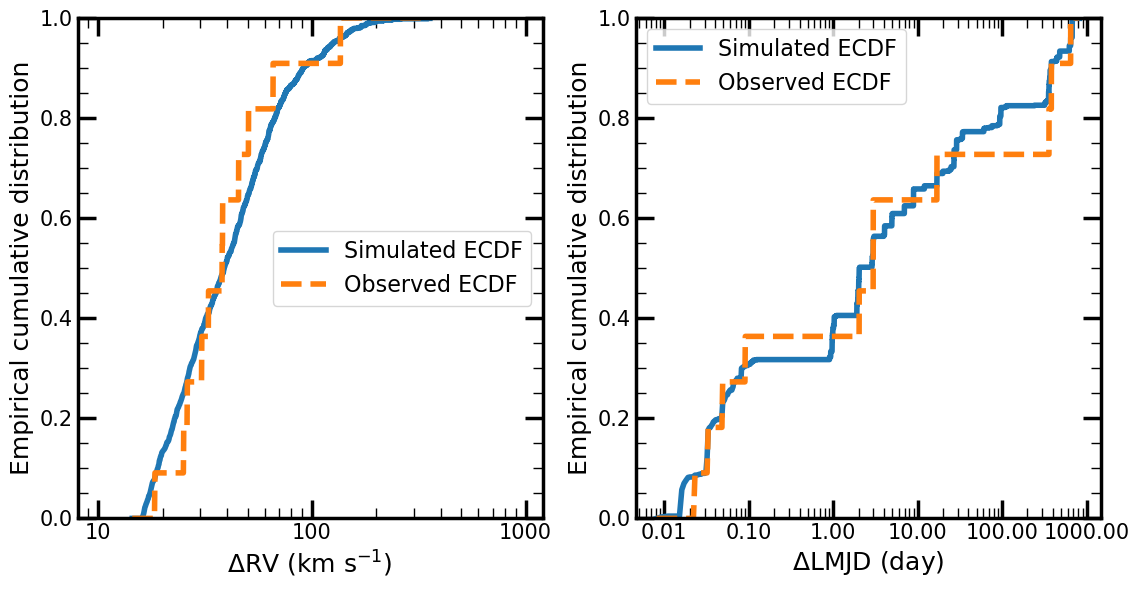}
\caption{Comparison between the simulated and observed ECDFs of $\Delta \rm RV$ (\textit{left panel}) and of $\Delta \rm LMJD$ (\textit{right panel}). The variables used to simulate the empirical cumulative distributions are $\pi=-5.7$, $\kappa=-3.6$ and $f_{\rm bin}=0.27$.
\label{fig:ECDF}}
\end{figure*}

Only 11 of the 203 stars with at least two observations satisfy the criteria of Equation (\ref{equ:criteria}) and are classified as spectroscopic binaries. The number of spectroscopic binaries is insufficient to fully reconstruct the parent distributions of $\Delta \rm RV$ and $\Delta \rm LMJD$. As a result, the constraints on the free variables provided by the GMF are weaker than those reported by \cite{Sana+.2013.A&A} within the range we explored. Additionally, the small values of the GMF suggest that even minor statistical fluctuations can lead to instability in its distribution within the three-dimensional variable space, particularly when the GMF is not well constrained. In our case, we computed the GMF multiple times to assess the stability of the results. The findings indicate that the location of the GMF maximum and its projections vary among different calculations due to statistical fluctuations. To determine the most stable location of the GMF maximum and improve the reliability of projections, we performed repeated GMF calculations at each grid point and used their average as the final GMF value. Since variable grid regions with the GMF values closer to the GMF maximum are more susceptible to statistical fluctuations, and considering the computational cost of the GMF calculations, we divided the variable grid into three regions according to their distance from the location of the GMF maximum. In order of increasing distance, the GMF at all grid points within these regions was computed 1400 times, 500 times, and 100 times, respectively. The regional projections of grid points with different calculation frequencies are shown in Figure \ref{fig:boundary}.

Figure \ref{fig:result} displays the projections of the GMF, which were obtained by analyzing the B-type runaway star sample. The maximum value of the GMF is obtained at $\pi=-5.7\pm 2.7$, $\kappa=-3.6\pm 1.8$, and an intrinsic binary fraction of $f_{\rm bin}=0.27\pm 0.08$. The quoted uncertainties are derived from the Monte Carlo simulations. Accordingly, on the basis of the observational sampling, the errors of RV estimates, and our best-fit parent distributions, we generated 100 synthetic data sets to assess the errors of the optimal variables. Due to the computational load of the error tests, we only calculated the GMF 500 times at each grid point within a narrow range around the optimal variables. Using our code, the GMFs of 31 out of the 100 data sets were well constrained. We ultimately used the larger of the two differences, $p_{50}-p_{16}$ and $p_{84}-p_{50}$, as the error estimate for each variable:
\begin{equation}
\sigma=\text{max}\{(p_{50}-p_{16}),(p_{84}-p_{50})\},
\label{equ:error}
\end{equation}
where $p_{16}$, $p_{50}$, and $p_{84}$ represent the 16th percentile, median, and 84th percentile, respectively, of the distribution of each optimal variable, derived from the analysis of the 31 well-constrained synthetic data sets. 

\section{Discussion} \label{sec:discussion}

The errors in our optimal variables are significantly larger than those in the optimal variables of \cite{Sana+.2013.A&A}, primarily because our sample includes only 11 detected spectroscopic binaries, resulting in a limited number of spectroscopic binaries being detectable in each synthetic data set. Thus, the observed distributions of $\Delta \rm RV$ and $\Delta \rm LMJD$ in the synthetic data sets are heavily influenced by randomness, leading to considerable differences between the optimal variables across the 31 synthetic data sets. This amplifies the differences between $p_{50}$ and $p_{16}$, and between $p_{84}$ and $p_{50}$.

Considering that the number of observations per star may affect the analysis results, and that only a small number of stars in our sample have at least four observations,  we repeated the binarity analysis on a subsample of 175 B-type stars with at least three observations. The resulting intrinsic binary fraction is $27\%$, consistent with that obtained from the larger sample of 203 B-type stars with at least two observations. This consistency supports the robustness of our results and indicates that the number of observations has little effect on the analysis within our sample.

\cite{Mason+.2009.AJ} and \cite{Aldoretta+.2015.AJ} reported observed spectroscopic binary fractions of $29\%$ and $28\%$ for O-type runaway stars, respectively. Given that the observed spectroscopic binary fraction represents a lower limit on the true binary fraction, the intrinsic binary fraction of O-type runaway stars is expected to be higher than these values. By comparison, our B-type runaway star sample exhibits a slightly lower intrinsic binary fraction of $27\%$. This discrepancy may primarily arise from two factors. First, O-type stars have a higher initial binary fraction than B-type stars, making binary systems composed of two O-type stars more likely to participate in the formation of runaway stars through both the BSS and the DES, thereby resulting in a greater number of O-type binary runaways. Second, compared to OB-type or BB-type binary systems, double O-type binaries typically have higher total masses and more compact orbits, which lead to greater binding energies, thereby making them less susceptible to disruption during SN explosions or dynamical interactions. As a result, they are more likely to survive as binary systems.

By combining the intrinsic binary fraction of our runaway star sample with the binary fractions of runaway stars formed via the BSS and the DES, as derived from binary population synthesis and $N$-body simulations, we can constrain the relative importance of the BSS and the DES. Assuming that runaway stars are formed solely through the BSS and the DES, and adopting a binary fraction of $20\%$ for the BSS runaway stars \citep{Portegies_Zwart.2000.ApJ} and $33\%$ for the DES runaway stars \citep{Perets+Subr.2012.ApJ}, the overall binary fraction of runaway stars can be expressed as 
\begin{equation}
    f_{\rm bin} = \frac{0.2X}{1+X}+\frac{0.33}{1+X},
\end{equation}
where $X$ is the ratio of the BSS to the DES. Therefore, the total binary fraction of $f_{\rm bin}=27\%$ in our runaway star sample suggests that the BSS and the DES play comparable roles in the Galactic B-type runaway population, with their relative contribution quantified as $X=0.86$. That is, approximately half of the runaway binary systems in our sample may have formed via the BSS. Given that the BSS is expected to produce HMXBs, we cross matched several HMXB catalogs to search for such systems. However, no corresponding sources were found. This suggests that our sample may contain undetected HMXBs or quiescent binary systems consisting of a B-type star and a neutron star or black hole.

Figure \ref{fig:ECDF} shows a comparison between the simulated and observed ECDFs of $\Delta \rm RV$ and of $\Delta \rm LMJD$. For the simulation based on the optimal variables, we estimated the overall binary detection probability for the runaway stars to be $f_{\rm bin}^{\rm obs}/f_{\rm bin}\simeq 20\%$. This implies that, after excluding the detected spectroscopic binaries, approximately $(f_{\rm bin}-f_{\rm bin}^{\rm obs})/(1-f_{\rm bin}^{\rm obs})\simeq 23\%$ of the remaining runaway stars are expected to be undetected binaries. The overall binary detection probability in our runaway star sample is substantially lower than the $69\%$ reported by \cite{Sana+.2013.A&A}. This discrepancy may result from differences in observational strategies. In \cite{Sana+.2013.A&A}, the time baseline for the vast majority of stars exceeds 300 days, whereas in our sample, $89\%$ of the stars have time baselines shorter than 100 days, as shown in Figure \ref{fig:observation}. Owing to the relatively short observational baselines, short-period binaries are easier to detect, while binaries with intermediate to long periods are more likely to be missed. Consequently, compared to the work of \cite{Sana+.2013.A&A}, our sample of detected spectroscopic binaries contains fewer systems with intermediate to long periods, which leads to a notable decrease in the overall binary detection probability. Moreover, the incomplete detection of mid-period binaries likely skews the observed ECDF of $\Delta \rm LMJD$ toward shorter time intervals, making it difficult to constrain the free variable $\pi$ and causing the optimal parent distribution of periods to deviate further from a uniform distribution.

\section{Conclusion}  \label{sec:conclusion}

This paper presents a multiplicity analysis of 203 Galactic B-type runaway stars from LAMOST DR8, cataloged by \cite{Guo+.2024.ApJS}. We found that the observed spectroscopic binary fraction in the runaway star sample is $5.4\%\pm 1.6\%$. Moreover, a fraction of $12.8\%\pm 2.3\%$ in the runaway stars exhibit significant RV variations, but with amplitudes less than $\rm 16~km~s^{-1}$. These stars may be potential binary systems or single stars with atmospheric activity.

By modeling the observational properties of the runaway star sample, we employed a Monte Carlo method to correct for observational biases and constrain its intrinsic multiplicity properties. Our simulation yielded an intrinsic binary fraction of $27\%\pm 8\%$. The period distribution that best fits the sample follows the power law of $f(P)\propto P^{-5.7}$, with $1\leq P\leq 1000$ days, indicating a preference for short-period binaries. Likewise, the most likely mass ratio distribution follows the power law of $f(q)\propto q^{-3.6}$, with $0.1\leq q\leq 1.0$, favoring less massive companions.

The intrinsic binary fraction we derived is marginally lower than the observed values reported by \cite{Mason+.2009.AJ} and \cite{Aldoretta+.2015.AJ}. Furthermore, our result for the binary fraction is substantially lower than that of B-type stars in young star clusters, suggesting that the formation of runaway stars is profoundly affected by both binary evolution and dynamical interaction, with a large portion of these stars originating from binary disruptions.

Combined with a binary fraction of $20\%$ for the BSS runaway stars from \cite{Portegies_Zwart.2000.ApJ} and $33\%$ for the DES runaway stars from \cite{Perets+Subr.2012.ApJ}, the intrinsic binary fraction of our runaway star sample implies that the contribution of the BSS to the Galactic B-type runaway star formation is at a similar level to that of the DES, with the ratio of the BSS to the DES being 0.86.

%% IMPORTANT! The old "\acknowledgment" command has be depreciated. It was
%% not robust enough to handle our new dual anonymous review requirements and
%% thus been replaced with the acknowledgment environment. If you try to 
%% compile with \acknowledgment you will get an error print to the screen
%% and in the compiled pdf.
%% 
%% Also note that the akcnowlodgment environment does not support long amounts of text. If you have a lot of people and institutions to acknowledge, do not use this command. Instead, create a new \section{Acknowledgments}.

\begin{acknowledgments}

This work was supported by the National Natural Science Foundation of China (grant Nos. 12288102, 12125303, 12090040/3, 12103064, 12403039, 12373036, 12473033, 12073070, and 12333008), the National Key R$\&$D Program of China (grant Nos. 2021YFA1600403/1 and 2021YFA1600400), the Strategic Priority Research Program of the Chinese Academy of Sciences (grant Nos. XDB1160201 and XDB1160000), the Natural Science Foundation of Yunnan Province (grant Nos. 202201BC070003 and 202001AW070007), the Yunnan Fundamental Research Project (grant Nos. 202401BC070007 and 202501CF070018), the International Centre of Supernovae, Yunnan Key Laboratory (grant No. 202302AN360001), the “Yunnan Revitalization Talent Support Program”--Science and Technology Champion Project (grant No. 202305AB350003), and the Yunnan Ten Thousand Talents Plan Young $\&$ Elite Talents Project.

\end{acknowledgments}

%% For this sample we use BibTeX plus aasjournals.bst to generate the
%% the bibliography. The sample631.bib file was populated from ADS. To
%% get the citations to show in the compiled file do the following:
%%
%% pdflatex sample631.tex
%% bibtext sample631
%% pdflatex sample631.tex
%% pdflatex sample631.tex

\bibliography{paper}{}
\bibliographystyle{aasjournal}

%% This command is needed to show the entire author+affiliation list when
%% the collaboration and author truncation commands are used.  It has to
%% go at the end of the manuscript.
%\allauthors

%% Include this line if you are using the \added, \replaced, \deleted
%% commands to see a summary list of all changes at the end of the article.
%\listofchanges

\end{document}